\documentclass[apjl]{emulateapj}
\newcommand{\etal}{{et al.~}}
\newcommand{\lta}{\la}

\newcommand{\kms}{\>{\rm km}\,{\rm s}^{-1}}

\newcommand{\Msun}{\>{\rm M_{\odot}}}
\newcommand{\msun}{\>{\rm M_{\odot}}}

\newcommand{\beq}{\begin{equation}}
\newcommand{\eeq}{\end{equation}}

\newcommand{\mpch}{\>h^{-1}{\rm {Mpc}}}

\newdimen\hssize
\newdimen\hdsize 
\shorttitle{Age Dependence of Halo Bias}
\shortauthors{Yang, Mo \& van den Bosch}

\begin{document}
            
%%%%%%%%%%%%%%%%%%%%%%%%%%%%%%%%%%%%%%%%%%%%%%%%%%%%%%%%%%%%%%%%%%%%%%%%%%

\title{Observational Evidence for an Age Dependence of Halo Bias}    

\author{Xiaohu Yang\altaffilmark{1,2}, H.J. Mo\altaffilmark{1} and
Frank C. van den Bosch\altaffilmark{3,4}}

\altaffiltext{1}{Department of Astronomy, University of Massachusetts,
      Amherst MA 01003-9305}
\altaffiltext{2}{present address: Shanghai Astronomical Observatory, 
      the Partner Group of MPA, Nandan Road 80, Shanghai 200030, China}
\altaffiltext{3}{Department of Physics, Swiss Federal Institute of
      Technology, ETH H\"onggerberg, CH-8093, Zurich, Switzerland} 
\altaffiltext{4}{present address: Max-Planck Institute for Astronomy, 
      D-69117 Heidelberg, Germany} 

%%%%%%%%%%%%%%%%%%%%%%%%%%%%%%%%%%%%%%%%%%%%%%%%%%%%%%%%%%%%%%%%%%%%%%%%%%

\begin{abstract}
  We study  the dependence  of the cross-correlation  between galaxies
  and galaxy groups on group properties.  Confirming previous results,
  we find that  the correlation strength is stronger  for more massive
  groups, in good agreement with  the expected mass dependence of halo
  bias. We also  find, however, that for groups of  the same mass, the
  correlation strength depends on the star formation rate (SFR) of the
  central galaxy: at  fixed mass, the bias of  galaxy groups decreases
  as  the  SFR of  the  central  galaxy  increases. We  discuss  these
  findings in  light of the recent  findings by Gao  \etal (2005) that
  halo  bias  depends on  halo  formation  time,  in that  halos  that
  assemble  earlier are  more strongly  biased.  We  also  discuss the
  implication  for galaxy formation,  and address  a possible  link to
  galaxy conformity,  the observed correlation  between the properties
  of satellite galaxies and those of their central galaxy.
\end{abstract}

%%%%%%%%%%%%%%%%%%%%%%%%%%%%%%%%%%%%%%%%%%%%%%%%%%%%%%%%%%%%%%%%%%%%%%%%%%


%%%%%%%%%%%%%%%%%%%%%%%%%%%%%%%%%%%%%%%%%%%%%%%%%%%%%%%%%%%%%%%%%%%%%%%%%%

\section{Introduction}
\label{sec:intro}

In  the  standard  cold   dark  matter  (CDM)  paradigm  of  structure
formation,  virialized CDM  halos are  considered to  be  the building
blocks of  the mass  distribution in the  Universe. The  properties of
dark matter halos, as well as their formation histories and clustering
properties,  have been studied  in great  detail using  both numerical
simulations as  well as analytical  approaches such as  the (extended)
Press Schechter formalism.  These studies have shown that halo bias is
mass dependent, in that more massive halos are more strongly clustered
(e.g., Mo \& White 1996; Seljak \& Warren 2004).  This mass dependence
of  the halo  bias  has played  a  crucial role  in understanding  the
correlation  function  of  both  dark  matter and  galaxies,  via  the
so-called halo model (e.g., Cooray \& Sheth 2002), the halo occupation
models  (e.g.,   Berlind  \&  Weinberg  2002),   and  the  conditional
luminosity function (e.g., Yang, Mo \& van den Bosch 2003).

Recently,  Gao  \etal  (2005)  used  a  very  large,  high  resolution
numerical  simulation   of  structure  formation   in  a  $\Lambda$CDM
cosmology  to reexamine  halo  bias.   They found  that  for halos  at
redshift $z=0$ with $M \lta 10^{13} h^{-1} \Msun$ the bias depends not
only on mass but also on the halo assembly time.  If the properties of
galaxies depend  on the assembly time  of their parent  halo, this may
have an important impact on the accuracy of halo occupation models and
the conditional  luminosity function, both of  which implicitly assume
that halo bias only depends on halo mass.
%In this  Letter we  present
%observational evidence that halo bias depends on the properties of the
%galaxies within each  halo.  Combined with the age  dependence of halo
%bias,  this provides a  strong indication  that (i)  galaxy properties
%indeed  are related  to  the epoch  of  halo assembly,  and (ii)  that
%current models for  halo occupation statistics need to  be improved to
%take proper account of this complex nature of halo bias.

\begin{deluxetable*}{cccccc}
\tabletypesize{\scriptsize}
\tablecaption{Galaxy and Group Samples.\label{tab:samp}}
\tablewidth{0pt}
\tablehead{
\colhead{Mass bin} & 
\colhead{$z$} & 
\colhead{$N_{\rm group}$} & 
\colhead{$N_{\rm galaxy}$} & 
\colhead{Mean $\eta_c$} &
\colhead{$b_{rel}$}
}

\startdata
$12.0 \le \log(M_{\rm h}/h^{-1}\Msun) \le 12.5$ & $[0.03~~0.11]$ & 7789  
& 14993 & 2.73/-0.28/-1.78/-2.85 & 0.76/0.93/0.95/1.18  \\
$12.5 \le \log(M_{\rm h}/h^{-1}\Msun) \le 13.0$ & $[0.03~~0.16]$ & 10984 
& 45139 & 2.06/-0.92/-2.19/-3.05 & 1.11/1.16/1.31/1.47  \\
$13.0 \le \log(M_{\rm h}/h^{-1}\Msun) \le 13.5$ & $[0.03~~0.16]$ & 3829  
& 45139 & 0.89/-1.70/-2.49/-3.16 & 1.30/1.36/1.50/1.82  \\
$13.5 \le \log(M_{\rm h}/h^{-1}\Msun) \le 14.0$ & $[0.03~~0.16]$ & 1117  
& 45139 & 0.07/-2.10/-2.68/-3.32 & 2.00/2.06/2.56/2.77  \\
\enddata

\tablecomments{Columns 1 and 2 list  the mass range and redshift limit
  of  each group  sample.  The  numbers of  groups and  galaxies (with
  $-21.5\le M_{b_J}-5\log  h\le -19.5$) in  each of these  samples are
  listed in columns 3 and 4. Groups in each sample are subdivided into
  four subsamples,  each containing one  quarter of the  total sample,
  according  to the  value of  the  spectral index,  $\eta_c$, of  the
  central  galaxy.  The  mean values  of  $\eta_c$  and  relative bias 
  $b_{rel}$ for  each of  these  subsamples  are indicated  in columns  
  5 and 6. }

\end{deluxetable*}

Since  dark matter  halos  are  thought to  mark  the locations  where
galaxies  form  and  reside,  a  promising  way  to  study  halo  bias
observationally is via the  clustering properties of galaxy groups. In
what  follows we  use a  very liberal  definition of  a  galaxy group,
including any system of galaxies  that belongs to the same dark matter
halo. This includes galaxy clusters as  well as halos that host only a
single galaxy (i.e., single  member galaxy `groups').  With the advent
of large galaxy redshift surveys, it is now possible to construct very
large group catalogues, which  allow an accurate, statistical study of
their  clustering  properties   (e.g.,  Merchan  \&  Zandivarez  2002;
Zandivarez  \etal 2003;  Yang  \etal 2005b,d;  Coil  \etal 2005).   In
addition,  these catalogues  allow for  a  detailed study  of how  the
properties of  the galaxy population  depend on the properties  of the
halo in which they reside  (Eke \etal 2004; Yang \etal 2005c; Weinmann
\etal 2005).

%%In  this Letter  we combine  both approaches  and investigate  how the
%%clustering  properties of  groups depend  on the  properties  of their
%%member   galaxies.  
In  this Letter  we  use the  galaxy-group cross-correlation  function
(hereafter GGCCF)  to study  its dependence on  the properties  of the
central galaxies  of the groups.  In  Section~\ref{sec_GC} we describe
the data  and the method used  for our analysis, the  results of which
are  presented  in  Section~\ref{sec_results}.   A discussion  of  the
implications of  our results for the  age dependence of  halo bias and
for galaxy conformity are discussed in Section~\ref{sec_discussion}.

\section{Data and analysis}
\label{sec_GC}

Our analysis  is based  on the group  catalogue of Yang  \etal (2005a;
hereafter YMBJ),  constructed from the 2-degree  Field Galaxy Redshift
Survey  (hereafter 2dFGRS;  Colless  \etal 2001).   This catalogue  is
constructed  with  a  new,  halo-based  group finder  which  has  been
optimized  to assign galaxies  into groups  according to  their common
dark  matter   halos.  
%%% The  group  finder  starts   with  an  assumed
%%% mass-to-light  ratio which  is used  to assign  each  potential group,
%%% identified with a friends-of-friends algorithm, with a tentative mass.
%%% This mass is then used  to estimate the radius and velocity dispersion
%%% of  the corresponding  dark matter  halo, which  in turn  are  used to
%%% determine  group memberships  in  redshift space.   This procedure  is
%%% applied iteratively until the  group memberships converge. 
Group masses  are estimated from  the ranking of group  luminosity, as
described in detail in Yang \etal (2005c).  As shown in Weinmann \etal
(2005), this  method yields masses  that are more accurate  than those
based  on  the  more  traditional  velocity dispersion  of  the  group
members.  However,  it requires knowledge  of the halo  mass function,
and  is   therefore  cosmology  dependent.   Throughout   we  adopt  a
$\Lambda$CDM   `concordance'   cosmology   with  $\Omega_m   =   0.3$,
$\Omega_{\Lambda}=0.7$, $h=0.7$  and $\sigma_8=0.9$.  We  use the form
given in Sheth, Mo \& Tormen (2001) for the halo mass function.

In Yang \etal (2005d) we used  this group catalogue to study the GGCCF
to quantify the spatial distribution of galaxies within CDM halos (see
also Collister  \& Lahav  2005; Paz \etal  2005).  In this  Letter, we
extend this analysis to study  how the GGCCF depends on the properties
of the central  group galaxies.  Motivated by the  current paradigm of
galaxy formation, we define the  central galaxy as the brightest group
member and we consider the  location of the central galaxy to coincide
with the  centre of  mass of the  group.  Note  that we use  the GGCCF
instead  of the auto-correlation  of groups,  because the  much larger
number of  galaxies (compared to the  number of groups)  allows a much
more accurate determination of the correlation power of the groups.

\begin{figure}
\plotone{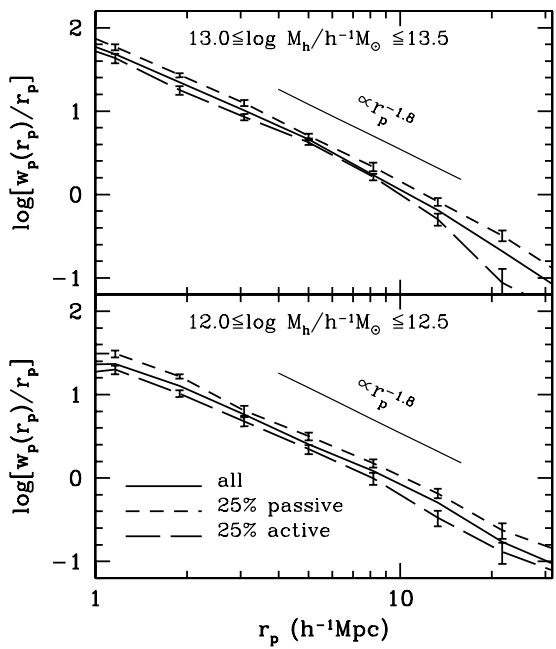}
\caption{Solid lines show the projected galaxy-group cross correlation 
  function (GGCCF). The upper and lower panels correspond to different
  group mass bins, as indicated.   For each mass bin, the short-dashed
  curves indicate  the GGCCF  for the subsample  with the  In this Letter we   use   the   galaxy-group
cross-correlation function (hereafter  GGCCF) to study its dependence
on  the  properties  of  the  central  galaxies  of  the  groups.   
In
Section~\ref{sec_GC} we describe the data  and the method used for our
analysis,    the     results    of    which     are    presented    in
Section~\ref{sec_results}.   A discussion of  the implications  of our
results for the age dependence  of halo bias and for galaxy conformity
are discussed in Section~\ref{sec_discussion}.
25\% lowest
  values of $\eta_c$ (i.e., the central galaxies with the most passive
  star  formation), while  the  long-dashed curves  correspond to  the
  subsample  with  the  25\%  highest  values of  $\eta_c$  (the  most
  actively  star forming  central galaxies).   Errorbars  indicate the
  1-$\sigma$ variance as obtained from 8 independent mocks.}
\label{fig:wrp}
\end{figure}

We split our group sample, at  fixed group mass, according to the star
formation rate (hereafter  SFR) of the central galaxy.  To this extent
we use the parameter $\eta$, which  is a linear combination of the two
most significant  principal components of  the 2dFGRS galaxy  spectra. 
As  shown   in  Madgwick  \etal  (2002),  $\eta$   follows  a  bimodal
distribution and is tightly correlated with the current SFR.  Galaxies
with $\eta\la  -1.4$ are mostly early-type galaxies  with passive star
formation, while  those with $\eta\ga  -1.4$ are mainly  actively star
forming,  late-type galaxies.  We  divide our  group sample  into four
mass bins.   Each of  these is further  subdivided into  4 equal-sized
subsamples according  to the value  of $\eta_c$ of the  central galaxy
(the subscript $c$  refers to the central galaxy).   Since for a given
group mass  the catalogue is only  complete out to  a certain redshift
limit (see  Yang \etal  2005d for details),  we restrict  the redshift
range to $0.03\le z  \le 0.11$ for the lowest mass bin,  and to $ 0.03
\le z  \le 0.16$ for  all other mass  bins.  For the galaxies,  we use
volume-limited   samples   with    absolute   magnitude   $-21.5   \le
M_{b_J}-5\log h \le -19.5$.  The various group and galaxy samples used
are listed in Table~\ref{tab:samp}.

In redshift space, the separation  between a group center and a galaxy
can  be split in  the components  perpendicular, $r_p$,  and parallel,
$\pi$, to  the line-of-sight. We  compute the GGCCF,  $ \xi(r_p,\pi)$,
using a  symmetrized version of  the Landy \& Szalay  (1993) estimator
(see Coil \etal 2005). The  random samples used for this estimator are
generated  taking  all  known  observational  selection  effects  into
account  (see Yang  \etal  2005d).  As  a  measure of  the real  space
correlation function we use the projected GGCCF, defined as
\begin{equation}
\label{project}
w_p(r_p) = \int_{-\infty}^{\infty} \xi(r_p,\pi) {\rm d}\pi \,.
\end{equation}
In practice, we  only integrate over the range  $\vert\pi\vert \leq 40
\mpch$, which suffices to capture all relevant correlation power.

\begin{figure}
\plotone{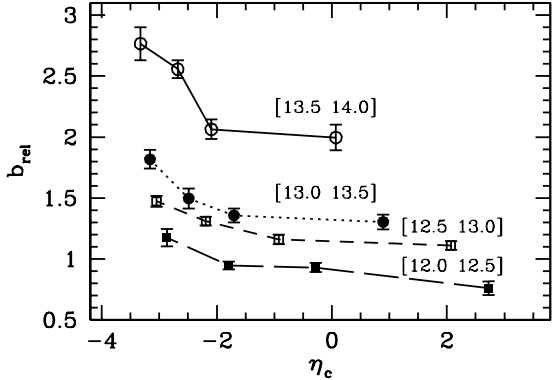}
\caption{The relative bias, $b_{\rm rel}$, for groups in 4 mass bins 
  as a function of $\eta_c$. Different symbols plus linestyles reflect
  different mass  bins, with the values in  square brackets indicating
  the range  of $\log(M_h/h^{-1}  \Msun)$.  For each  mass bin,  the 4
  subsamples all contain  the same number of groups,  and the value of
  $\eta_c$  plotted  is  the   average  value  for  the  corresponding
  subsample.   The offset between  the curves  for the  different mass
  bins reflects the  mass dependence of the halo  bias.  For groups of
  the same mass, however,  these is also a $\eta_c$-dependence: groups
  in which the central galaxy has a more passive star formation (i.e.,
  a lower value of $\eta_c$) are more strongly clustered.}
\label{fig:bias_eta}
\end{figure}

\section{Results}
\label{sec_results}

Fig.~\ref{fig:wrp} shows  the projected  GGCCF for groups  with masses
$12.0  \le \log(M_{\rm h}/h^{-1}  \Msun) \le  12.5$ (lower  panel) and
$13.0 \le \log(M_{\rm h}/h^{-1} \Msun) \le 13.5$ (upper panel).  Solid
lines indicate the  results for {\it all} groups  in the corresponding
mass bin,  while short-dashed and long-dashed lines  correspond to the
subsamples  with  the 25\%  lowest  and  highest  values of  $\eta_c$,
respectively.   Clearly, groups  with  a more  passive central  galaxy
(i.e.,  a lower  value of  $\eta_c$) have  a  higher cross-correlation
amplitude.   Since the  same galaxies  are used  in the  estimation of
these GGCCFs, the relative amplitude  between them is a measure of the
relative clustering bias of the different groups.

For $r_p\ga 3\mpch$, all the GGCCFs are well described by a power law,
$w_p(r_p)\propto r_p^{-0.8}$.  As discussed  in Yang \etal (2005d), on
these scales  the GGCCF  is dominated by  the `2-halo' term,  which is
determined  by the halo-halo  correlation. In  order to  determine the
relative bias of different groups,  we fit all the projected GGCCFs in
the range $3\le r_p \le 30\mpch$ with the power-law,
\begin{equation}
\label{est_bias}
w_p(r_p) = A \, b_{\rm rel} \, r_p^{-0.8}\,,
\end{equation}
where we set $A=50$ so that $b_{\rm rel}\approx 1$ for the full sample
of all groups with $12.0 \le  \log(M_{\rm h}/h^{-1} \Msun) \le 12.5 $. 
The   values   of  $b_{\rm   rel}$   thus   obtained   are  shown   in
Fig~\ref{fig:bias_eta} for groups with  different values of $\eta_c$. 
The errorbars here  are estimated from the 1-$\sigma$  scatter among 8
mock samples (see YMBJ).  As  one can see, the relative bias increases
strongly with group mass, as  expected from the mass dependence of the
halo  bias (e.g.,  Mo \&  White 1996;  Seljak \&  Warren 2004)  and is
consistent  with  earlier observational  results  (Padilla 2004;  Yang
\etal 2005b,d).   For a given  mass bin, there  is a clear  trend that
groups  with a  smaller $\eta_c$  (i.e., with  a more  passive central
galaxy) have higher $b_{\rm rel}$.  The ratio of $b_{\rm rel}$ between
the quarters with  the smallest and the highest  values of $\eta_c$ is
about $1.4$-$1.6$ for each of the four mass bins.

\begin{figure}
\plotone{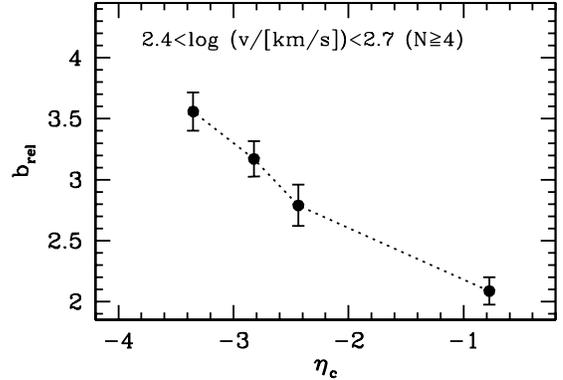}
\caption{The same as Fig.~\ref{fig:bias_eta}, but here 
the mass of each group is estimated from the velocity dispersion 
of its member galaxies.}
\label{fig:bias_vel}
\end{figure}

The group masses used above have been determined assuming a one-to-one
relation between halo mass and the group luminosity in the $b_J$-band.
One might  argue that  halo mass is  more closely associated  with the
total {\it  stellar} mass of the  galaxies, rather than  with the blue
light. If this is indeed the case, our method of assigning halo masses
may introduce an artificial $\eta_c$-dependence:  a halo with a lot of
recent  star formation  (i.e., with  a  high $\eta_c$  value) will  be
overly luminous  in the $b_J$-band,  so that we will  overestimate its
mass.  Since lower mass haloes are less strongly biased, this could in
principle result  in a false detection of  $\eta_c$-dependence of halo
bias. For relatively  rich groups we can test  this using the velocity
dispersion of the  member galaxies as a dynamical  mass estimator.  In
Fig.\,\ref{fig:bias_vel} we  show the relative  bias as a  function of
$\eta_c$  for  groups with  4  member or  more  that  have a  velocity
dispersion in the  range of $250 \kms$ to $500  \kms$.  Here again the
relative  bias increases  as $\eta_c$  decreases, suggesting  that the
$\eta_c$-dependence of  $b_{\rm rel}$ is not simply  due to systematic
errors in our halo masses.

\begin{figure}
\plotone{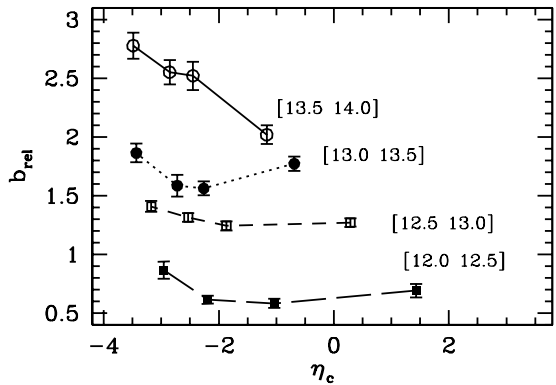}
\caption{The same as Fig.~\ref{fig:bias_eta}, but here 
the mass of each group is based on the total stellar mass  
of its member galaxies.}
\label{fig:bias_stellar}
\end{figure}

For poor groups  where the velocity dispersion is  not a reliable mass
estimator, we  test our  results using the  total {\it  stellar} mass,
instead of the total $b_J$-band luminosity, to determine group masses.
Using the  9200 galaxies in the  2dFGRS that are also  included in the
Sloan Digital Sky Survey (SDSS;  NYU-VAGC Blanton \etal 2005), we find
a  mean  relation  between  the  stellar mass-to-light  ratio  $M_*  /
L_{b_J}$ and $\eta$ given by
\begin{equation}
\label{ml}
{\rm log}\left[ {M_*/L_{b_J}} \right] = -0.088 \eta + 0.587\,.
\end{equation}
The  stellar masses  for these  galaxies  are obtained  from the  SDSS
spectra as described in Kauffmann \etal (2003). Using eq.(\ref{ml}) we
compute  $M_*$ for  all galaxies  in our  2dFGRS group  catalogue, and
estimate the  halo masses  using the total  stellar mass of  all group
members. The  resulting relations  between $b_{\rm rel}$  and $\eta_c$
are shown in Fig.\,\ref{fig:bias_stellar}.  Note that there is still a
significant  $\eta_c$-dependence at  fixed  halo mass.   Nevertheless,
there  are some  differences  with  respect to  the  results shown  in
Fig.~\ref{fig:bias_eta}. In particular, for  the bins with the highest
$\eta_c$ values, the relative bias has increased with respect to using
the  group luminosity to  assign the  halo masses.   This owes  to the
effect discussed  above, and is  particularly pronounced for  the mass
bin $\log(M_h/h^{-1}\msun)=[13.0-13.5]$.  For  less massive haloes the
effect  is weaker,  simply because  for  haloes with  $M \lta  10^{13}
h^{-1} \Msun$  the halo bias  only depends weakly  on $M$, so  that an
error in  halo mass  has only  a small effect.   For the  most massive
haloes, the  effect is also  much weaker, basically because  for these
systems the luminosity of the  central galaxy is only a small fraction
of the total group luminosity. We  wish to stress that since it is not
{\it a priori} clear whether halo mass is more tightly correlated with
stellar mass or  with $b_J$-band luminosity, it is  not clear which of
the  results (Fig.~\ref{fig:bias_eta}  or Fig.~\ref{fig:bias_stellar})
are the more accurate.  Overall, however, our results indicate a clear
$\eta_c$  dependence of  halo  bias. An  accurate  measurement of  the
absolute  strength of  this effect,  however, requires  a  more robust
determination of halo masses.

\section{Discussion}
\label{sec_discussion}

%We have shown  that the clustering strength of  galaxy groups does not
%only  depend on the  group (halo)  mass, but  also on  the SFR  of the
%central galaxy.   The mass dependence  is well understood in  terms of
%the mass  dependence of the  halo bias, and  has been noticed  before. 
%The dependence on the properties  of the central galaxy, however, is a
%new result.  It is intriguing to link this result to that of Gao \etal
%(2005), who, using numerical simulations, found that halos of the same
%mass that assemble earlier are more strongly clustered, i.e. halo bias
%is age dependent.  Our results  therefore seem to suggest that central
%galaxies residing in halos that assembled earlier are more passive (in
%terms of their current SFR) than central galaxies in halos of the same
%mass, but which assembled later.

Galaxies are thought to form in CDM halos, and it is generally assumed
that galaxy properties are only  determined by the properties of their
host halo (e.g., mass,  angular momentum, formation history, etc.). In
particular, in the `standard'  picture, adopted in all semi-analytical
models  of galaxy  formation, the  morphology of  a central  galaxy is
related to the epoch of  the last major merger: halos that experienced
their last major merger more recently (i.e., that assembled later) are
more likely to host an  early type (passive) central galaxy.  Based on
the results of Gao \etal (2005), one would expect halos with a passive
central galaxy  to be less strongly  clustered than halos  of the same
mass, but  with a late type  (active) central galaxy,  contrary to the
results presented here. In order  to explain the SFR dependence of the
halo bias, one needs a mechanism  that shuts off the star formation of
the  central galaxy  earlier in  a  halo that  assembles earlier.  For
example, if the time of the last major merger also signals the time at
which star  formation is  terminated, a redder  central galaxy  may be
produced by an  earlier major merger.  The age  dependence of the halo
bias would then be in qualitative agreement with the results presented
here. Interestingly, a similar truncation of star formation seems also
required in order  to explain the bright end  of the galaxy luminosity
function, and may be related to AGN feedback (e.g., Benson \etal 2003;
Granato  \etal 2004;  Nagashima \etal  2004; Croton  \etal  2005).  It
remains  to be  seen  whether semi-analytical  models  that take  such
feedback  processes into  account  can indeed  explain the  clustering
dependencies presented here.

It is  also interesting to link  the results presented  here to galaxy
conformity. As  shown in  Weinmann \etal (2005),  halos with  an early
type central galaxy have a significantly larger fraction of early type
satellites than a halo of the  same mass, but with a late type central
galaxy.  The results presented  here, therefore,  suggest a  bias that
depends not  only on the properties  of the {\it  central} galaxy, but
also on those  of the {\it entire} galaxy population  of the group.  A
halo that assembled earlier will have typically accreted its satellite
population  earlier.   It is  generally  assumed  that  once a  galaxy
becomes a satellite  galaxy of a bigger system,  its star formation is
truncated, either because the galaxy loses its hot gas supply (Larson,
Tinsley \&  Caldwell 1980), or because ram  pressure stripping removes
the  cold gas supply.   This would  suggest that  halos with  a larger
fraction of  early type  (passive) satellites assembled  earlier.  The
age dependence of the halo bias then implies that these systems should
be more strongly clustered,  in qualitative agreement with the results
presented here.

However, a  number of open questions  remain. First of  all, Gao \etal
(2005) only detected  an age  dependence of halo  bias for  halos with
masses below  the characteristic  non-linear mass scale  $M^{*} \simeq
10^{13} h^{-1}  \Msun$, whereas  we find that  the more  massive halos
also  reveal an $\eta_c$-dependence.   Second, it  is unclear  how the
halo formation  time, defined in Gao  \etal (2005) as the  time when a
halo assembles half of it mass,  is related to the age of the galaxies
that  form in  the  halo.  Numerical  simulations and  semi-analytical
models are required to investigate these issues in detail.
  
%%%%%%%%%%%%%%%%%
% Ackowledgements
%%%%%%%%%%%%%%%%%
                                                                           
\acknowledgments We  thank Simon White and Asantha  Cooray for helpful
comments.  FvdB acknowledges lively  discussions on galaxy groups with
Alison Coil,  Risa Wechsler and  Andreas Berlind.  XY is  supported by
the  {\it One  Hundred  Talents}  project of  the  Chinese Academy  of
Sciences.
                                                                               
%%%%%%%%%%%%%%%%%
% Bibliography
%%%%%%%%%%%%%%%%%

\clearpage

\end{document}